# About Standpoints for Surmounting Reversibility and Recurrence Paradoxes in Theoretical Thermodynamics

Yoshihiro Nakato*

*Department of Materials Engineering Science, Graduate School of Engineering Science, Osaka University, Toyonaka, Osaka 560-8531, Japan (presently retired)*

The reversibility and recurrence paradoxes are key issues that have been left unsolved in researches on the foundation of thermodynamics since the 19th century. This article shows that (1) the reversibility paradox can be overcome if we pay attention to observed quantities, which represent net results of the motion of a reversible character in the non-observable area and (2) the recurrence paradox can be surmounted if we look at an isolated system from the inside, in which there is no system with discrete energy levels. These arguments confirm the validity of the reported conclusion that the origin of the increase of entropy in an isolated system is the basic quality of microscopic particles of spontaneously spreading into every possible spatial area.

Recently, I reported that a new reasonable definition can be given to the entropy of an isolated system if we pay attention to the basic quality of microscopic particles such as electrons and atomic nuclei of spontaneously spreading into every possible spatial area.[1] Based on this result, I also reported that we can directly deduce thermodynamic laws from quantum mechanics with no assumption and statistical treatment.[2] These results have an outstanding feature in that the entropy of an isolated system is directly connected with the basic quality of microscopic particles and the meaning of it is easy to understand. Now, such great success has been achieved by considering an isolated system from some new standpoints, independently of current trends of researches on the foundation of thermodynamics,[3] but the standpoints were not explained in detail in reported papers.[1,2] In particular, I did not explain that the new standpoints allow us to surmount severe paradoxes in theoretical thermodynamics, called the reversibility paradox and the recurrence paradox, which have been left unsolved since the 19th century.[3] The aim of this article is to explain this point clearly.

The reversibility paradox arises from the fact that mechanical laws such as Newton's law of motion and the Schrödinger equation, which describe the motion of constituent particles in an isolated system, have a reversible character with respect to the reversal of time, contrary to irreversible thermodynamic laws such as the law of increase of entropy. For example, Newton's law of motion, $m\,dx^2/dt^2 = F$, does not change the form by the alteration of $t$ to $-t$. Similarly, the Schrödinger equation for an isolated system, for which the Hamiltonian operator $\mathcal{H}(q)$ is independent of time,

$$i\hbar(\partial/\partial t)\,\Psi(q,t) = \mathcal{H}(q)\,\Psi(q,t), \qquad (1)$$

also has a reversible character with respect to the reversal of time, because the complex conjugate of Eq. (1),

$$-i\hbar(\partial/\partial t)\,\Psi(q,t)^* = \mathcal{H}(q)\,\Psi(q,t)^*, \qquad (2)$$

has the same form as an equation in which $t$ in Eq. (1) is converted to $-t$.

This problem can be solved if we notice that only observed quantities are real to us.

Indeed, the Schrödinger equation for an isolated system has a reversible character, as mentioned above, but both Eqs. (1) and (2) express the motion of microscopic particles in the non-observable area. On the other hand, observable quantities such as the probability of finding a system, given by $\Psi(q,t)^*\Psi(q,t)\mathrm{d}\tau$, and an expectation value of a physical quantity $\Omega(q)$, given by $\langle\Omega(t)\rangle \equiv \langle\Psi(q,t)^*\Omega(q)\Psi(q,t)\rangle$, are always expressed in the form including both a wave function $\Psi(q,t)$ in Eq. (1) and its complex conjugate $\Psi(q,t)^*$ in Eq. (2), indicating that an observed quantity represents a *net result* of the motion of microscopic particles in the forward and the backward directions in the non-observable area. Therefore, an observed quantity has no longer any connection with a reversible character of the Schrödinger equation.

The recurrence paradox arises from Poincaré's recurrence theorem, which says that an isolated system repeatedly returns to the initial state in the course of its time evolution within an error of regulation of initial conditions. Quantum mechanics has a similar theorem, called the quantum recurrence theorem.[4] Such a recurrent character of mechanical laws is also incompatible with irreversible thermodynamic laws. This problem can be solved if we look at an isolated system from the inside. In this case, we only look at individual particles or sub-systems interacting with each other. Thus, the quantum recurrence theorem, which only applies to an isolated system with discreate energy levels, does not apply because individual particles or sub-systems interacting with each other have no discrete energy levels.

In conclusion, both the reversibility and recurrence paradoxes can be overcome by adopting standpoints of (1) distinguishing observable quantities and non-observable ones and (2) looking at an isolated system from the inside. The first standpoint arises from the general argument that originally, we can only observe "results" of the motion of particles and the motion itself is in the non-observable area, or in other words, a moving or changing thing is impossible to directly observe. In fact, for a moving particle, we can only say, "A particle is here at an instant and is there at a next instant", but these words describe nothing about motion itself. When we say, "At an instant, a particle is here", a particle stops here at this instant. The same holds at a next instant. A particle moves between an instant and a next instant but no description is given to the behavior of a particle in this period. Thus, the above words only describe a "result" of motion. Differential calculus in mathematics also uses the same method of description except that the period between an instant and a next instant is made infinitesimal. The second standpoint arises from the general argument that originally, an isolated system is in principle impossible to look at from the outside, because we cannot look at any system without interaction. Nobody can look at the universe from the outside. Some other merits of looking at an isolated system from the inside are discussed in Ref. 2.

*A correction to Ref. 1*: The group velocity of a wave packet is described as $\hbar k/m$ in Eq. (3b) of Ref. 1 but this is wrong and corrected to $\hbar k_0/m$.